# Your Reviews Replicate You: LLM-Based Agents as Customer Digital Twins for Conjoint Analysis

Bin Xuan

Department of Data Science Seoul National University of Science and Technology Seoul, Republic of Korea xuanbin159@seoultech.ac.kr

Jungmin Hwang *

Department of Data Science Seoul National University of Science and Technology Seoul, Republic of Korea

inextro@ds.seoultech.ac.kr

Hakyeon Lee*

Department of Industrial Engineering Seoul National University of Science and Technology Seoul, Republic of Korea

hylee@seoultech.ac.kr

Conjoint analysis is a cornerstone of market research for estimating consumer preferences; however, traditional methods face persistent challenges regarding time, cost, and respondent fatigue. To address these limitations, this study proposes a framework that utilizes large language model (LLM)-based "customer digital twins (CDT)" as virtual respondents. We identified active users within the Reddit community and aggregated their comprehensive review histories to construct individualized vector databases. By integrating retrieval-augmented generation (RAG) with prompt engineering, this study developed customer agents capable of dynamically retrieving and reasoning upon their specific past preferences and constraints. These customer agents, called CDTs, performed pairwise comparison tasks on product profiles generated via fractional factorial design, and the resulting choice data was analyzed to estimate part-worth utilities by logistic regression. Empirical validation demonstrates that these CDTs predict the preferences of actual users with 87.73% accuracy. Furthermore, a case study on the computer monitor category successfully quantified trade-offs between attributes such as panel type and resolution, deriving preference structures consistent with market realities. Ultimately, this study contributes to marketing research by presenting a scalable alternative that significantly improves both agility and cost-efficiency to traditional methods.

---

* Corresponding authors.

# 1 INTRODUCTION

The success of new product development depends significantly on the ability to accurately understand customer preferences and incorporate them into product design. Regardless of how innovative a firm's technology may be, market success cannot be assured if the product fails to deliver the value that customers genuinely seek [5]. In this context, conjoint analysis has established itself as a core method in marketing research since the 1970s. Conjoint analysis has contributed to deriving optimal product configurations by quantitatively measuring how consumers evaluate various product attributes and perform trade-offs [24]. Since the pioneering work of Green and Srinivasan [12], conjoint analysis has been applied across diverse product categories ranging from consumer goods to industrial products over several decades, and its validity and utility have been demonstrated through numerous studies. This method is particularly distinguished from traditional surveys and focus group interviews in its capacity to estimate not only directly stated preferences but also preference structures that emerge indirectly through the choice process [21].

However, traditional conjoint analysis possesses inherent limitations in the data collection process. Recruiting large-scale respondent panels and administering surveys require considerable time and cost, which constitutes a significant constraint in product development environments where rapid decision-making is essential. Furthermore, due to the nature of conjoint surveys that require respondents to repeatedly evaluate complex attribute combinations, respondents tend to exhibit diminished attention or rely on simplified heuristics as tasks progress [12]. Various improved methods have been proposed to address these issues. Choice-based conjoint analysis (CBC) enhanced ecological validity by simulating more realistic decision-making situations where respondents select one option among multiple alternatives, while adaptive conjoint analysis (ACA) increases survey efficiency and reduces respondent fatigue by analyzing previous response patterns in real-time and dynamically adjusting subsequent questions [12, 15]. Nevertheless, improved conjoint analysis methods including CBC and ACA still share the fundamental constraint of dependence on human respondents [6]. This challenge is particularly pronounced for target segments with limited accessibility, such as B2B markets or expert populations, where securing appropriate respondents itself poses a substantial difficulty [13]. This factor constrains the representativeness and reliability of analytical results and leads to conjoint analysis being conducted primarily as a one-time exercise at specific stages of product development [12]. Consequently, these methodological constraints are not well aligned with iterative and agile product improvement processes.

Meanwhile, recent advances in large language model (LLM) technology present new possibilities for overcoming the fundamental constraint of dependence on human respondents. State-of-the-art LLMs such as GPT-4 have demonstrated capabilities in understanding complex contexts, performing consistent judgments from the perspective of specific roles, and exhibiting human-like reasoning patterns [4, 23]. These capabilities are grounded in human linguistic expressions, values, and decision-making patterns learned through vast textual data [3, 14]. Particularly noteworthy prior research has empirically demonstrated that when actual consumer interview data is provided to an LLM to train it on that consumer's preference system, the LLM exhibits choice patterns similar to those of that consumer [32]. This finding suggests that LLMs can function as digital twins capable of replicating the preferences of real individuals, beyond merely simulating non-existent virtual personas. These prior research achievements suggest that if customer digital twins (CDTs) can substitute for human respondents in conjoint analysis, the methodological constraints discussed earlier could be substantially alleviated. In particular, such digital twins significantly reduce the time and cost required for recruiting large-scale respondent panels. They can also provide consistent responses without fatigue or attention degradation and can be utilized repeatedly as needed to ensure alignment with agile product improvement processes.

Against this background, this study proposes a framework that substitutes LLM-based CDT for human respondents who have traditionally performed conjoint analysis. The proposed framework consists of two main stages. First, active users related to a specific product category are identified from online communities, and their review data is collected to create CDTs integrated with a retrieval-augmented generation (RAG) system. Subsequently, the generated digital twins perform conjoint choice tasks for product profiles, and the collected data is statistically analyzed to derive part-worth utilities for each product attribute. To validate the proposed framework, this study empirically analyzes whether review data-based CDTs exhibit choice patterns similar to those of actual users and further explores the practical applicability of part-worth utilities derived through this framework.

## 2 RELATED WORK

### 2.1 Conjoint Analysis

Conjoint analysis is a method that quantitatively analyzes the preference structures that consumers form by simultaneously considering multiple attributes of products or services [12]. The theoretical foundation of this method can be traced to Lancaster's consumer theory, in which he argued that consumers derive utility not from products themselves but from combinations of attributes that products provide [18]. From this perspective, a product is understood as a bundle of multiple attributes, and consumer choice is made in a direction that maximizes the sum of part-worth utilities provided by each attribute [27]. Green and Rao [11] introduced conjoint analysis to the marketing field upon this theoretical foundation, and subsequently Green and Srinivasan [13] established it as a standardized method through comprehensive review papers. The analytical framework they completed enabled the calculation of attribute-specific values, which had previously been discussed only theoretically, as concrete numerical figures termed part-worth utilities. These quantified utility values clearly demonstrate the relative importance of each attribute, thereby serving as the core basis for firms to make practical marketing decisions such as deriving optimal product configurations or analyzing price sensitivity.

Early traditional conjoint analysis adopted the full-profile rating method, presenting respondents with multiple product profiles and requiring them to assign preference scores to each [12]. While this method had the advantages of being intuitive and easy to analyze, it faced two fundamental limitations. First, the rating task itself did not accurately reflect actual purchasing situations where consumers choose one option among alternatives rather than evaluating each option independently. Second, the exponential increase in the number of profiles according to combinations of attributes and levels imposed excessive cognitive burden on respondents, resulting in decreased response reliability as the survey progressed [13]. Subsequent methodological developments addressed these distinct limitations through different approaches. CBC, proposed by Louviere and Woodworth, enhanced ecological validity by simulating more realistic decision-making situations where respondents select one option among multiple alternatives, and enabled analysis through discrete choice models. ACA, developed by Sawtooth Software, focused on improving survey efficiency by dynamically adjusting subsequent questions based on respondents' previous responses [16]. Further refinements such as Adaptive Choice-Based Conjoint (ACBC) integrated the ecological validity of CBC with the efficiency of ACA [17], while Hierarchical Bayesian estimation improved individual-level preference estimation with limited observations per respondent [19].

Nevertheless, all conjoint analysis methods share a fundamental constraint in that they presuppose the participation of actual human respondents. This leads to difficulties in respondent recruitment and data quality degradation as respondent fatigue accumulates during survey processes [8, 29]. Furthermore, high per-respondent costs make it difficult to conduct large-scale or iterative analyses [7]. Although online panel services have made respondent recruitment easier, securing appropriate respondents remains difficult in segments requiring specialized expertise or in markets with limited accessibility [9, 30]. These constraints demand a transition to a new data collection paradigm.

In this context, LLMs that have internalized vast human data have emerged as a promising alternative capable of complementing the limitations of traditional surveys. If LLMs can validly simulate the preference systems of actual consumers, this suggests the possibility of iterative experimentation without physical constraints and agile market analysis. Therefore, to utilize LLMs as reliable CDTs rather than mere text generation tools, theoretical examination must precede such application. The following section examines recent trends in LLM-based agents as virtual respondent research and discusses the validity of integrating this into conjoint analysis.

### 2.2 LLM-Based Agents as Virtual Respondents

Attempts to utilize LLMs as substitutes for or complements to human respondents are being concretized through concepts such as "silicon sampling" or "synthetic respondents" [3, 14]. This approach is premised on the notion that LLMs, in the process of learning massive textual data, have internalized not only human knowledge but also social norms, attitudes, and nuanced linguistic variations [2]. If the model can reproduce the distinctive preference systems and judgment criteria of specific human groups or individuals, this implies that researchers can obtain an efficient tool for validating hypotheses without physical constraints [4, 23]. This section examines existing research by categorizing it into two types according to the implementation method and depth of personalization of LLM-based agents as virtual respondents, "group-level" and "individual-level."

Group-level methods aim to imitate the average responses of groups defined by demographic or psychological characteristics through prompt engineering. This approach is grounded in the premise that LLMs have internalized the attitudinal tendencies and value systems of different social groups through training on massive textual data [2, 28]. Empirical studies have demonstrated that appropriately prompted LLMs can reproduce group-specific response patterns in various survey and experimental tasks [1, 3, 14].

Individual-level methods do not rely on the average characteristics of groups but rather aim to reproduce the unique preference systems and decision-making patterns of individual consumers. The core concern lies in capturing the preference heterogeneity that inevitably exists among individuals even when they belong to the same demographic group [31]. The key challenge in achieving this goal is how to convey sufficiently rich individual background information to the LLM so that the model can infer the unique latent preference structure of that individual [22, 31].

However, conveying individual information solely through prompts faces evident capacity limitations, making it difficult to fully utilize the substantial historical data that individuals accumulate through daily consumption activities [22, 31]. RAG technology provides an effective solution to this problem by storing individual historical information in an external vector database and dynamically retrieving relevant content according to the current task context during inference, thereby overcoming the constraints of prompt length [22, 31]. Research findings indicate that when LLMs are provided with sufficient background information through such systems, the model can infer the latent preference structure of specific individuals and form stable response patterns reflecting their decision-making tendencies [26].

When examining these technical possibilities within the context of conjoint analysis, individual-level methods prove more aligned with the methodological requirements of conjoint analysis. Since Green and Srinivasan established its foundation, the fundamental purpose of conjoint analysis has been to estimate part-worth utilities of individual consumers and identify preference heterogeneity among them [10]. If only group-level methods are employed, multiple agents assigned identical demographic characteristics may exhibit overly similar response patterns, leading to underestimation of preference differences observed in actual markets. Furthermore, conjoint analysis requires respondents to repeatedly evaluate multiple product configurations based on consistent value judgment criteria, necessitating that LLM-based agents as virtual respondents maintain stable preference systems rooted in authentic consumption experiences.

In light of this, this study proposes an integrated framework that develops LLM-based agents to construct CDTs capable of substituting for human respondents using RAG technology. While previous approaches have constructed individual profiles using interview transcripts or survey responses [26, 32], such methods still require deliberate data collection from human participants, thereby not fully resolving the cost and accessibility constraints inherent in traditional conjoint analysis. This study instead leverages product reviews accumulated in online communities, which already exist at scale without additional collection efforts and capture consumers' authentic usage experiences expressed in naturalistic settings. This approach enables the construction of individual preference models for diverse consumer segments, including those with specialized expertise or limited accessibility, directly addressing the sample acquisition bottlenecks identified in conventional methods. The following section will elaborate on the specific architecture of this framework and the implementation process of each component.

## 3 PROPOSED APPROACH

### 3.1 Framework

A methodological research gap exists between the constraints of respondent acquisition in traditional conjoint analysis and the potential of LLMs to simulate personalized human behavior. This study proposes a framework that approximates actual customers' preference systems through CDTs designed by combining prompt engineering and RAG and applies insights derived from these digital twins to product development decision-making.

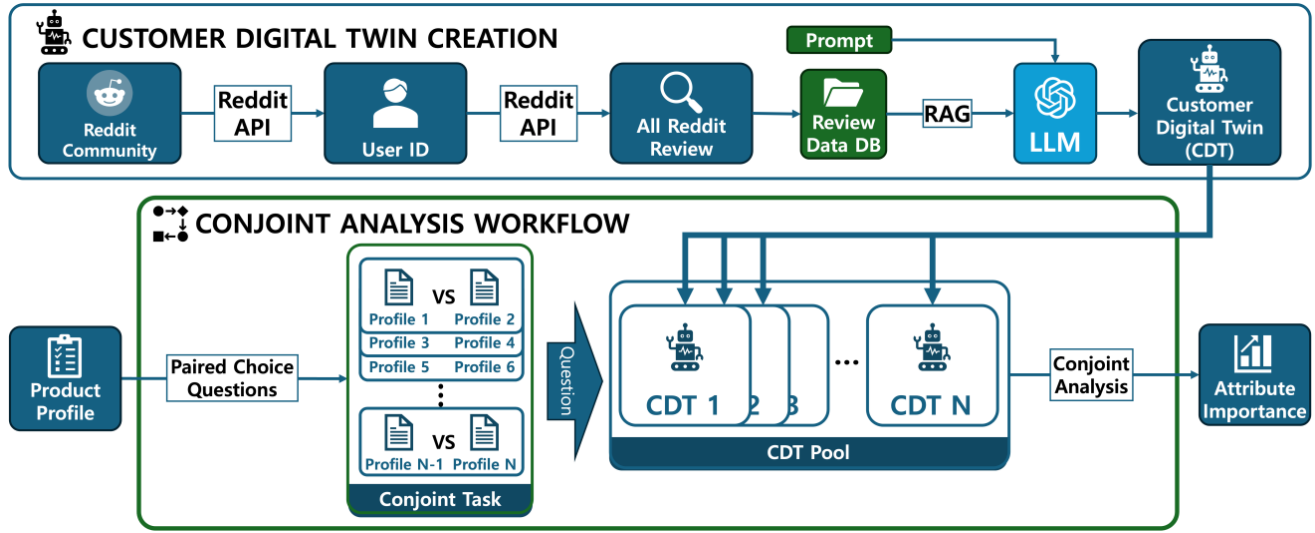

Figure 1: Framework Overview

Figure 1 presents the overall structure of the proposed framework. The framework consists of two main components. The upper component, Customer Digital Twin Creation, is the process of collecting actual customers' review data and generating CDTs that reflect individual customers' preference systems through the combination of RAG and prompts. The CDTs generated in this manner function as virtual customer proxies that serve as respondents in conjoint analysis. The lower component, Conjoint Analysis Workflow, is the process whereby the generated CDTs perform pairwise comparison tasks on product profiles, and the results are analyzed to derive attribute importance for product concept improvement. The two components are interconnected such that the CDTs created in the upper component persist as a CDT pool, and during conjoint analysis tasks, each CDT draws on its review history through RAG to produce pairwise choices.

### 3.2 Customer Digital Twin Creation

The Customer Digital Twin Creation presented in the upper portion of Figure 1 is the process of creating CDTs to be utilized as respondents in the Conjoint Analysis Workflow. This process consists of a sequential flow comprising actual customer data collection, vector database construction, and CDT generation, and produces CDTs that reflect specific customers' preference systems and judgment criteria using prompts and RAG.

The starting point of CDT generation is securing data that enables understanding of actual customers' preference systems. This study utilizes Reddit communities as the data source, as this platform hosts diverse product-focused subreddits where users voluntarily share detailed usage experiences and evaluative opinions within communities of shared interest. First, users active in subreddits related to specific product categories are identified using the Reddit API, and user IDs are extracted. Subsequently, all reviews and comments written by each user across the entire Reddit platform are comprehensively collected. The rationale for not limiting data collection to product-specific subreddits is to capture latent contextual signals, such as occupation, lifestyle, and typical usage patterns, that support preference inference when explicit preferences are unavailable.

The collected review data is structured and stored in the review data DB and integrated with the RAG system. Each customer's review data is converted into vector embeddings and stored in individual vector databases, and the RAG architecture enables CDTs to dynamically retrieve and reference past reviews with high semantic relevance to the current context when performing conjoint choice tasks. For instance, when eliciting preferences for a specific product attribute, opinions that the user previously expressed regarding that attribute are preferentially retrieved and provided to the LLM. This grounds the CDT's preference expression by supplying contextual information directly relevant to the current choice rather than merely enumerating past data.

Another core element of CDT generation is the design of prompts that induce the LLM to behave from the perspective of a specific customer. The prompt explicitly defines the role the CDT must perform and the factors to consider during preference elicitation. In particular, the prompt in this study is designed so that the CDT comprehensively analyzes the information provided through RAG and the user's overall tendencies, enabling reasonable preference inference with the user's latent preferences even in domains where direct data is absent. The specific prompt composition used in this study is presented in Appendix A.

As presented in Figure 1, each CDT combines prompt-defined role instructions with RAG-retrieved review data to form a preference system grounded in the LLM's reasoning capabilities. These CDTs serve as respondents in the Conjoint Analysis Workflow, with the CDT pool representing heterogeneous customer segments of the target market.

### 3.3 Conjoint Analysis Workflow

The Conjoint Analysis Workflow presented in the lower portion of Figure 1 represents the process of performing systematic conjoint choice tasks using the CDTs generated in Section 3.2 and deriving attribute importance and part-worth utilities from the results. This system maintains the theoretical foundation of existing conjoint analysis while substituting CDTs for respondents. The starting point of the conjoint analysis is defining the attributes of the product to be evaluated. The product is decomposed into multiple attributes, and each attribute has several levels. The selection of attributes and levels is performed reflecting product development objectives and market context and is limited to technically feasible combinations. Product profiles are generated from the defined attribute combinations, with each profile representing a complete product configuration in which a specific level is assigned to every attribute.

The generated profiles are structured in the form of paired choice questions. The adoption of the pairwise comparison method in this study is attributable to its demonstrated advantages in LLM-based preference measurement. Preference elicitation through pairwise comparison shows higher concordance with human judgment than direct scoring methods [20], enhances accuracy by simplifying complex judgments into relative comparisons of single pairs [25], and mitigates bias problems that persist even with calibration techniques in direct scoring approaches [20]. As presented in the conjoint task area of Figure 1, the CDT pool independently performs the conjoint tasks, with each CDT comparing two presented product profiles and selecting the preferred alternative. When CDTs perform tasks, the RAG mechanism constructed in the upper pipeline retrieves and provides the corresponding customer's past opinions related to the attributes being compared, thereby ensuring that CDT choices are based on preference patterns rather than arbitrary judgment.

While the pairwise comparison method is advantageous for eliciting stable and human-like preference expression from LLM-based CDTs, the number of required comparisons increases exponentially with attributes and levels; this study addresses this through orthogonal design that enables efficient estimation with a manageable number of comparisons. The collected choice data is statistically analyzed in the conjoint analysis stage to estimate part-worth utilities of each attribute level, which are then converted into attribute importance quantifying each attribute's contribution to overall product preference. The derived attribute importance enables the identification of product configurations aligned with target market preferences by computing expected utilities through the summation of part-worth utilities. The proposed framework supports iterative and rapid customer preference elicitation by achieving personalization and stable preference expression through CDT design that combines prompt engineering and RAG technology.

## 4 PRELIMINARY VALIDATION

### 4.1 Validation Method

Prior to applying the proposed framework to practical market analysis, preliminary validation of how accurately the constructed CDTs simulate actual users' preferences is essential. To this end, this study selected the Reddit 'r/Monitors' community as the data source for empirical analysis, as this community contains substantial user-generated content discussing product attributes and preferences. Using the Reddit API, the top 200 active users with high posting and commenting frequency were identified, and up to 1,000 posts and comments were collected for each user, yielding approximately 200,000 records in total. Following the methodology described in Section 3.2, CDTs were constructed for all 200 users by integrating their review data into individual vector databases. To construct the ground truth dataset for validation, reviews containing direct comparisons of two attribute levels with explicit preference statements were extracted, yielding 163 comparison reviews written by 85 distinct users.

The validation methodology is based on the revealed preference principle. If a user directly compares two levels of a specific attribute in a review and expresses explicit preference for one of them, this can be regarded as an externalization of that user's actual preference system. As illustrated in Figure 2, when a user expresses preference for IPS over QD-OLED panels with specific reasons, this reveals that user's part-worth utility structure for the panel type attribute. From this statement, a binary choice question was constructed asking which panel type the user prefers between IPS and QD-OLED, with IPS as the ground truth (e.g., Question: IPS vs. QD-OLED, Ground Truth: IPS). This study extracted such comparison statements and converted them into ground truth data to evaluate the predictive validity of CDTs.

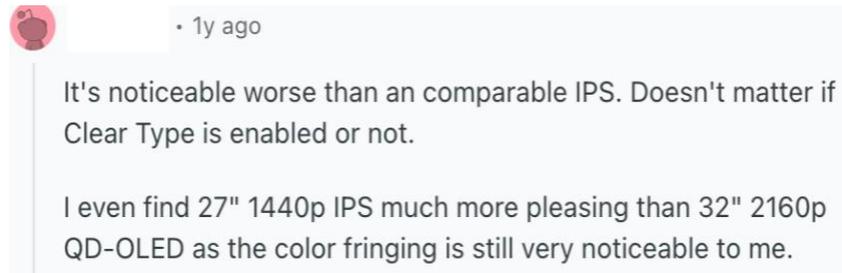

Figure 2: Example of ground truth construction from user comparison statement

A core consideration in the validation design was the prevention of information leakage. If an CDT references the review containing the comparison expression itself, this would merely amount to extracting the preference explicitly stated in that review; therefore, this study applied strict temporal separation, providing CDTs only with review history prior to the point when the relevant review was written through the RAG system. This design enables validation of the consistency and generalizability of the preference system by ensuring that CDTs perform selections based solely on preference patterns learned from past reviews. Validation proceeded by constructing a RAG database based on each user's past history for each of the 163 comparison reviews, requesting the CDT to select one of the two attribute levels, and then evaluating concordance by contrasting this with the preference expressed by the actual user.

### 4.2 Validation Results

The validation results for the 163 comparison tasks revealed that CDTs achieved an accuracy of 87.73%. Of the total 163 cases, 149 yielded correct answers and 14 yielded incorrect answers. This substantially exceeds the expected accuracy of 50% for random selection, suggesting that review data-based CDTs are meaningfully learning the preference systems for attribute levels of the corresponding users.

Analysis of correct cases reveals that CDTs accurately predicted user-specific preferences rather than attribute levels generally considered superior. Figure 3 illustrates that one user expressed a preference for 1440p over 4K resolution in a review, stating that "4k is overrated" and that "1440p looks amazing, 4k for me isn't necessary". The CDT accurately identified this preference pattern from past reviews and correctly selected 1440p in the validation task. Similar patterns were observed across different attribute comparisons, including cases where CDTs correctly identified users' preferences for specific panel types or refresh rates based on concerns expressed in their review history, such as usability issues or perceived diminishing returns at higher specifications.

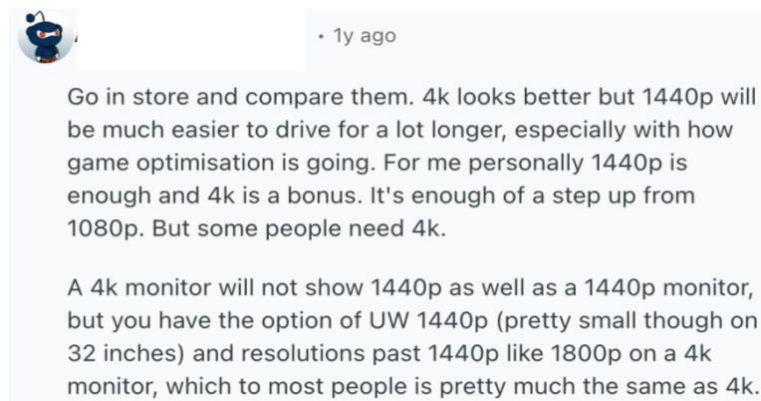

Figure 3: Example of CDT preference prediction based on user review history

These results suggest that the proposed framework reproduces individual users' preferences with a high degree of accuracy. This level of performance indicates that the framework can be effectively applied to conjoint analysis, where capturing individual-level preference structures is essential.

# 5 CASE STUDY

## 5.1 Research Design

While the preceding validation confirmed that CDTs can effectively reproduce actual users' preferences at the individual-level, demonstrating the framework's utility for practical marketing decisions requires application to a realistic product choice scenario. To this end, this study conducted a case study targeting computer monitor products. Monitors are a product category well-suited for conjoint analysis due to the existence of various technical attributes such as screen size, resolution, refresh rate, and panel type, with clear trade-offs among attributes. Furthermore, monitors are appropriate for observing heterogeneity among customer segments because preferences for attributes differ according to usage purposes such as gaming, professional creative work, and office tasks. The availability of abundant review data necessary for CDT generation, given that consumers actively exchange opinions about technical attributes of monitors in online communities, was also considered.

Product attributes and levels were derived through consultation with a product planning manager at a monitor specialized company to ensure relevance to actual product development decisions and technical feasibility. Table 1 presents the 5 selected attributes and their respective levels.

The full combination of 5 attributes with 2 levels each generates 32 ($2^5$) product profiles. However, executing a full factorial design necessitates an excessive number of comparisons, resulting in significant experimental inefficiency. To address this limitation, this study employed an orthogonal design methodology utilizing an $L_{16}$ orthogonal array (equivalent to a $2^{5-1}$ fractional factorial design). This methodology allowed for the selection of 16 balanced profiles, enabling the independent estimation of main effects while minimizing correlations among attributes. Furthermore, a foldover method was utilized to construct the choice tasks, in which each of the 16 orthogonal profiles was paired with its complementary (mirror) profile. Consequently, this design resulted in 16 distinct pairwise comparison questionnaires.

Table 1: Definition of Product Attributes and Levels

| Attribute | Level 1 | Level 2 |
|---|---|---|
| Screen Size | 27-inch | 34-inch |
| Aspect Ratio | 16:9 (Standard) | 21:9 (Ultrawide) |
| Panel Type | OLED Pro | IPS Black |
| Refresh Rate | 120Hz | 240Hz |
| Resolution Class | 4K-class | 8K-class |

This case study was conducted using CDTs constructed from the same 200 active users described in Section 4. Because Reddit users maintain consistent accounts while operating anonymously, tracking opinions and preference patterns of specific users over extended periods is facilitated. Following the methodology discussed in Section 3.2, the vector database construction stage for each CDT involved comprehensive collection of all posts and comments written by the corresponding user across the entire Reddit platform, not limited to monitor-related subreddits. This was intended to comprehensively reflect lifestyle contexts that could directly influence monitor preferences, such as game genres, video editing activities, and coding environments. To this end, up to 1,000 recently written posts and comments were collected for each user and loaded into the RAG system's vector database, thereby implementing a structure enabling CDTs to reason based on users' holistic context rather than fragmentary product reviews.

System implementation was performed using the LangGraph framework, which enables the definition and orchestration of LLM-based workflows in graph structures. In this study, LangGraph was used to integrate the RAG system and the conjoint query process into a single pipeline. Each CDT stored the corresponding user's review data in a vector database and retrieved relevant reviews when pairwise comparison questions were issued, supplying them as context to the LLM. The OpenAI GPT-5.1 model was selected as the LLM API due to its stable performance in handling long-context inputs and its compatibility with the LangGraph-based agent architecture. Based on this configuration, 200 CDTs each responded to 16 pairwise comparison questions, and attribute importance and part-worth utilities were estimated from the resulting choice data.

## 5.2 Results

The 3,200 pairwise comparison response data collected from 200 CDTs were analyzed through logistic regression analysis. The model's pseudo R² was 0.182, which corresponds to a generally acceptable level of explanatory power in conjoint analysis. The log-likelihood ratio test result confirmed the statistical significance of the overall model with a p-value less than 0.001. Table 2 presents the part-worth utility estimates and statistical significance of each attribute.

Table 2: Logistic Regression Analysis Results

| Attribute | Coefficient | Standard error | Z-value | P-value |
|---|---|---|---|---|
| Intercept | 0.795 | 0.044 | 17.978 | <0.001 |
| Screen Size (34-inch) | 0.484 | 0.043 | 11.242 | <0.001 |
| Aspect Ratio (21:9) | 0.033 | 0.042 | 0.776 | 0.438 |
| Panel Type (IPS Black) | -0.774 | 0.044 | -17.537 | <0.001 |
| Refresh Rate (240Hz) | 0.376 | 0.043 | 8.797 | <0.001 |
| Resolution (8K-class) | -0.688 | 0.044 | -15.711 | <0.001 |

Note: Pseudo R² = 0.182, Log-Likelihood = -1697.5, N = 3,200

The analysis results indicated that four attributes excluding aspect ratio exerted statistically significant effects at the p<0.001 level. When relative importance was calculated based on the absolute values of each attribute's coefficients, panel type exhibited the highest importance at 32.9%, followed by resolution at 29.2%, screen size at 20.6%, refresh rate at 16.0%, and aspect ratio at 1.4%. Table 3 presents the relative importance by attribute.

Table 3: Relative Importance by Attribute

| Attribute | Utility | Relative importance |
|---|---|---|
| Panel Type | 0.774 | 32.9% |
| Resolution | 0.688 | 29.2% |
| Screen Size | 0.484 | 20.6% |
| Refresh Rate | 0.376 | 16.0% |
| Aspect Ratio | 0.033 | 1.4% |

The optimal product profile can be derived by combining the preferred levels from each attribute. Table 4 presents the most preferred and least preferred product combinations calculated based on part-worth utility values.

Table 4: Optimal and Least Preferred Product Profiles

| Ranking | Screen size | Aspect ratio | Panel type | Refresh rate | Resolution | Total utility |
|---|---|---|---|---|---|---|
| Highest preference | 34-inch | 21:9 | OLED Pro | 240Hz | 4K-class | 1.688 |
| Least preference | 27-inch | 16:9 | IPS Black | 120Hz | 8K-class | -0.667 |

The optimal product profile emerged as the combination of 34-inch screen size, 21:9 ultrawide aspect ratio, OLED Pro panel, 240Hz refresh rate, and 4K-class resolution. The total utility value of this combination is 1.688, representing the sum of part-worth utilities of the preferred levels from each attribute. Conversely, the least preferred combination was 27-inch, 16:9, IPS Black, 120Hz, and 8K-class, showing a total utility value of -0.667. The utility difference between the two profiles is 2.355, demonstrating that the combination of attribute levels has a substantial impact on product preference. The strong preference for OLED Pro over IPS Black reflects perceived superiority in image quality. The preference for 4K-class over 8K-class can be attributed to current content scarcity and high hardware requirements for 8K. Screen size and refresh rate preferences align with demand for immersion and smoother visuals. Aspect ratio showed no statistical significance (p=0.438), suggesting that ultrawide benefits are offset by compatibility concerns. In summary, customers consider panel type and resolution as most important in monitor purchasing decisions, with these two attributes accounting for 62.1% of total importance. Screen size and refresh rate show moderate levels of importance, while aspect ratio exhibits relatively lower influence. These results suggest that product planning should prioritize attention to panel technology and resolution selection, and that aspect ratio can be utilized as a strategic differentiation element for targeting specific segments.

## 6 DISCUSSION

Using CDTs as substitutes for human respondents in conjoint analysis offers practical benefits, including lower cost and faster analysis. At the same time, LLMs do not behave in the same way as human respondents. As a result, conjoint analysis designs developed for humans cannot be directly applied without adjustment. This highlights the need to explore design principles that reflect the unique characteristics of LLMs. This section discusses considerations for designing LLM-based conjoint systems from three perspectives identified during the research process, question design, model operation, and personalization.

Regarding question design, this study compared three question formats. These formats were pairwise comparison, ranking-based questions, and choice-based conjoint (CBC). The comparison examined which format was suitable for preference elicitation in an LLM-based setting. Using identical product attributes and profiles while varying only the question format, the analysis of CDT response characteristics confirmed that each method exhibits different characteristics in CDT environments. The ranking-based method, despite its advantage of generating an entire preference structure with a single question, exhibited low prompt compliance and low response consistency due to increased prompt length. The CBC method provides a selection structure similar to actual purchasing situations but exhibited lower response consistency with certain attributes being undervalued due to selection pressure. In contrast, the pairwise comparison method achieved very high prompt compliance and very high response consistency, as its simple task structure of comparing only two profiles substantially reduced the possibility of hallucination or rule collapse. The limitation of question number proliferation in exhaustive pairwise comparison was considerably alleviated through the application of orthogonal design discussed in Section 3.3. This approach enables efficient estimation with a manageable number of comparisons. Regarding model operation, an experiment was conducted performing identical conjoint tasks with three settings of 0, 0.5, and 1 for the temperature parameter to examine its effect on LLM responses. Experiments with temperature settings of 0, 0.5, and 1 showed consistent choice patterns, indicating that the binary nature of pairwise comparison renders responses insensitive to this parameter and enhances system stability.

Regarding personalization, the core of this study is that CDTs represent specific customers rather than general virtual personas. This distinction is enabled by the use of a RAG system. In the proposed framework, RAG operates in two ways. When CDTs perform pairwise comparison tasks and past reviews related to the attributes currently being compared are retrieved, CDTs utilize this content as direct evidence for making judgments. Even when related reviews are not retrieved, CDTs do not rely on the LLM's general knowledge but instead retrieve multiple recent reviews of the corresponding customer and generate responses by comprehensively considering preferences, usage environments, and personality characteristics appearing therein. This design enables CDTs to reason based on the overall context of the corresponding customer even when direct mentions of specific attributes are absent. To verify the effectiveness of the RAG system, tests were conducted comparing responses between CDTs with RAG applied and CDTs without RAG applied for identical customer data, and clear response differences were confirmed between the two conditions. CDTs without RAG applied exhibited similar response patterns regardless of individual customer characteristics, whereas CDTs with RAG applied expressed differentiated preferences based on each customer's review data. These results empirically demonstrate that the RAG system plays a core role in CDT personalization.

## 7 CONCLUSION

This study proposed an LLM-based CDT framework utilizing online review data to address the cost and time constraints associated with respondent recruitment in traditional conjoint analysis. Extending the approach of prior research based on interview data, a methodology was developed that constructs digital twins having learned individual customers' preference systems from publicly accessible Reddit community user data and utilizes these as respondents for conjoint analysis. The proposed framework was designed with a structure that generates CDTs by integrating RAG systems and LLMs, and these CDTs perform pairwise comparison conjoint tasks to derive attribute importance and part-worth utilities. Empirical analysis results demonstrated that CDTs constructed through this framework predicted actual users' preferences with approximately 87.73% accuracy, thereby proving that they simulate users' value judgment criteria at a meaningful level beyond simple text generation. The case study targeting computer monitor products also derived that panel type and resolution are core attributes in product selection and presented interpretable preference structures by calculating statistically significant part-worth utilities. This confirms the methodological possibility of transforming unstructured text data in the form of online reviews into quantitative preference data. The study holds academic significance in that it overcomes the limitations of existing group-level persona research and implements personalized CDTs by reflecting individual consumers' heterogeneous preference structures through RAG technology.

Nevertheless, this study possesses several limitations. The most fundamental limitation is that validation comparing CDT choices directly with actual Reddit users whose data was utilized for CDT generation was not conducted. Although CDTs exhibited preference structures with internal consistency and personalization effects from RAG application were confirmed, whether the CDTs accurately reproduce the actual preferences of those users requires separate validation research. Furthermore, because only data collected from the single platform of Reddit was utilized, there are limitations in generalizing the results, and Reddit users may differ from general consumer populations in demographic characteristics or technology involvement. Whether the empirical results from the technology product category of monitors can be equally applied to other product categories such as fashion, food, and services must be confirmed through additional research. Additionally, due to the rapid development of LLM technology, the results of this study may be limited to specific models and time points. Despite these limitations, this study holds significance in that it explored the possibility of conjoint analysis utilizing LLM-based CDTs and derived design principles for practical application, and it is anticipated that the generalizability of the framework will be verified through future extended research across diverse platforms and product categories.

## A  Prompts

### ROLE & PERSONA

You are the Reddit user '{user_id}'.

The content provided below represents your OWN past memories, reviews, and opinions.

You must simulate this specific user's preference logic, writing style, and decision-making criteria.

### TASK

Your task is to choose between two options based **strictly** on your retrieved memories.

If your memories do not explicitly mention these specific options, infer the most likely choice based on your past preferences (e.g., brand loyalty, feature priorities, price sensitivity).

### OPTIONS TO COMPARE

- Option A: {option_a}
- Option B: {option_b}

### YOUR MEMORIES (Context)

{retrieved_memories}

### OUTPUT FORMAT INSTRUCTION

1. Analyze the memories to determine which option aligns better with your past self.
2. You MUST return the result in a valid JSON format.
3. Do not include any markdown formatting (like ```json) or additional text. Just the raw JSON string.

### REQUIRED JSON OUPUT

{{
  "choice": "A"  // or "B"
}}